\documentclass[12pt]{article}
\usepackage{amsmath}
\usepackage{amssymb}
\usepackage{graphicx}
\usepackage{axodraw}
\usepackage{cite}
\usepackage{url}
\usepackage[small]{caption}
\begin{document}
\baselineskip 0.6cm

\def\bra#1{\langle #1 |}
\def\ket#1{| #1 \rangle}
\def\inner#1#2{\langle #1 | #2 \rangle}
\def\app#1#2{%
  \mathrel{%
    \setbox0=\hbox{$#1\sim$}%
    \setbox2=\hbox{%
      \rlap{\hbox{$#1\propto$}}%
      \lower1.1\ht0\box0%
    }%
    \raise0.25\ht2\box2%
  }%
}
\def\approxprop{\mathpalette\app\relax}

\begin{titlepage}

\begin{flushright}
UCB-PTH-15/02 \\
\end{flushright}

\vskip 1.2cm

\begin{center}
{\Large \bf A Note on Boltzmann Brains}

\vskip 0.7cm

{\large Yasunori Nomura}

\vskip 0.5cm

{\it Berkeley Center for Theoretical Physics, Department of Physics,\\
 University of California, Berkeley, CA 94720, USA}

\vskip 0.2cm

{\it Theoretical Physics Group, Lawrence Berkeley National Laboratory,
 CA 94720, USA}

\vskip 0.2cm

{\it Kavli Institute for the Physics and Mathematics of the Universe (WPI),\\ 
 Todai Institutes for Advanced Study, University of Tokyo, 
 Kashiwa 277-8583, Japan}

\vskip 0.8cm

\abstract{Understanding the observed arrow of time is equivalent, under 
 general assumptions, to explaining why Boltzmann brains do not overwhelm 
 ordinary observers.  It is usually thought that this provides a condition 
 on the decay rate of every cosmologically accessible de~Sitter vacuum, 
 and that this condition is determined by the production rate of Boltzmann 
 brains calculated using semiclassical theory built on each such vacuum. 
 We argue, based on a recently developed picture of microscopic quantum 
 gravitational degrees of freedom, that this thinking needs to be modified. 
 In particular, depending on the structure of the fundamental theory, 
 the decay rate of a de~Sitter vacuum may not have to satisfy any 
 condition except for the one imposed by the Poincar\'{e} recurrence. 
 The framework discussed here also addresses the question of whether 
 a Minkowski vacuum may produce Boltzmann brains.}

\end{center}
\end{titlepage}

\section{Introduction}
\label{sec:intro}

At first sight, the fact that we observe that time flows only in one 
direction may seem mysterious, given that the fundamental laws of physics 
are invariant under reversing the orientation of time.%
\footnote{The operation discussed here is not what is called the time 
 reversal $T$ in quantum field theory, which we know is broken in 
 nature.  It corresponds to $CPT$ in the standard language of quantum 
 field theory.}
Upon careful consideration, however, one notices that the problem 
is not the unidirectional nature {\it per~se}.  As discussed in 
Refs.~\cite{Aguirre:2011ac,Nomura:2011rb}, given any final state 
$\ket{f}$ whose coarse-grained entropy is lower than the initial state 
$\ket{i}$, the evolution history is overwhelmingly dominated by the 
$CPT$ conjugate of the standard (entropy increasing) process $\ket{\bar{f}} 
\rightarrow \ket{\bar{\imath}}$.  This implies that a physical observer, 
who is necessarily a part of the whole system, sees virtually always, 
i.e.\ with an overwhelmingly high probability, that time flows from 
the ``past'' (in which correlations of the observer with the rest of 
the system are smaller) to the ``future'' (in which the correlations 
are larger).

The problem of the arrow of time, therefore, is not to understand its 
unidirectional nature, but to explain why physical predictions are 
(probabilistically) dominated by what we observe in our universe, 
i.e.\ a flow from a very low coarse-grained entropy state to a slightly 
higher entropy state.  In particular, it requires the understanding 
of the following facts:
\begin{itemize}
\item
At least one set of states representing our observations, which are 
mutually related by time evolution spanning the observation time, are 
realized in the quantum state representing the whole universe/multiverse. 
(Here and below we adopt the Schr\"{o}dinger picture.)  This is the 
case despite the fact that these states have very low coarse-grained 
entropies.%
\footnote{Because of the Hamiltonian constraint, the full 
 universe/multiverse state is expected to be static, i.e.\ not 
 to depend on any time parameter~\cite{DeWitt:1967yk,Nomura:2012zb}. 
 We may, however, talk about effective time evolution if we focus 
 on branches of the whole universe/multiverse state, since they are 
 not (necessarily) invariant under the action of the time evolution 
 operator $e^{-i H \tau}$.  This is the picture we adopt in this paper. 
 Note that this time evolution still does not have to be the same 
 as ``physical time evolution'' defined through correlations among 
 physical subsystems, e.g., as in Ref.~\cite{Page:1983uc}.  In the 
 static-state picture, the statement here is phrased such that the 
 state of the universe/multiverse contains components representing 
 our observations despite the fact that they are not generic in 
 the relevant Hilbert space.}
\item
The answer to a physical question, which may always be asked in 
the form of a conditional probability~\cite{Nomura:2011dt}, must 
be determined by the class of low coarse-grained entropy states 
described above.  In particular, the probability should not (always) 
be dominated by the states in which the unconditioned part of the 
system has the highest coarse-grained entropies.
\end{itemize}
These elements comprise (essentially) the well-known Boltzmann brain 
problem~\cite{Dyson:2002pf,Albrecht:2002uz}.  The problem of the 
arrow of time is thus {\it equivalent} to the Boltzmann brain 
problem~\cite{Nomura:2012zb} under (rather general) assumptions 
that went into the line of argument given above.

Any realistic cosmology must accommodate the two facts listed above. 
Is it trivial to do so?  In a seminal paper~\cite{Dyson:2002pf}, Dyson, 
Kleban, and Susskind pointed out that it is not.  In particular, they 
considered a de~Sitter vacuum representing our own universe and argued 
that if it lives long enough, thermal fluctuations in de~Sitter space 
lead to Boltzmann brains observing chaotic worlds, who overwhelm ordinary, 
ordered observers like us.  If true, this would give an upper bound 
on the lifetime of our universe which is much stronger than that needed 
to avoid the Poincar\'{e} recurrence  (barring the possibility that 
the observed vacuum energy relaxes into a zero or negative value in 
the future).  In this paper we argue that this consideration needs 
to be modified, based on the picture of the microscopic structure of 
quantum gravity advanced recently~\cite{Nomura:2014voa} to address 
the black hole information problem~\cite{Hawking:1976ra,Almheiri:2012rt}. 
We discuss implications of this modification for our own universe 
and the eternally inflating multiverse.  We also discuss implications 
of the framework for the possibility~\cite{Page:2005ur} of Boltzmann 
brain production in a Minkowski vacuum.

\section{de~Sitter Space in Quantum Gravity}
\label{sec:dS}

We first extend the discussion of Ref.~\cite{Nomura:2014voa}, which 
mainly focused on a system with a black hole, to de~Sitter space. 
In cosmology, de~Sitter space appears as a meta-stable state in the 
middle of the evolution of the universe/multiverse, and in this sense 
it is similar to a spacetime with a dynamically formed black hole. 
Indeed, string theory suggests that there is no absolutely stable 
de~Sitter vacuum in full quantum gravity; it must decay, at least, 
before the Poincar\'{e} recurrence time~\cite{Kachru:2003aw}.  This 
implies that what we call de~Sitter space cannot be an eigenstate 
of energy (at least in this context).

Consider a semiclassical de~Sitter space with Hubble radius $\alpha$. 
(We focus on 4-dimensional spacetime for simplicity, but the extension 
to other dimensions is straightforward.)  Following the complementarity 
hypothesis~\cite{Susskind:1993if}, and in particular its implementation 
in Refs.~\cite{Nomura:2011rb,Nomura:2011dt}, we adopt a ``local 
description,'' in which quantum states represent physical configurations 
on equal-time hypersurfaces foliating the causal patch associated 
with a freely falling frame.  We assume that the timescale for the 
evolution of microstates representing the de~Sitter space is of order 
$\varDelta t \approx \alpha$, where $t$ is the proper time measured 
at the spatial origin, $p_0$, of the reference frame.  The uncertainty 
principle then implies that a state representing this space must 
involve a superposition of energy eigenstates with a spread of order 
$\varDelta E \approx 1/\alpha$.  Associating this energy with the 
vacuum energy density $\rho_\Lambda$ integrated over the Hubble volume, 
$E \approx O(\rho_\Lambda \alpha^3) \approx O(\alpha/l_{\rm P}^2)$, 
this spread is translated into $\varDelta \alpha \approx 
O(l_{\rm P}^2/\alpha)$, where $l_{\rm P}$ is the Planck length.

How many different independent ways are there to superpose the energy 
eigenstates to arrive at the semiclassical de~Sitter space described 
above?  As in the black hole case, we assume that the Gibbons-Hawking 
entropy~\cite{Gibbons:1977mu}
\begin{equation}
  S_{\rm GH} = \frac{{\cal A}}{4 l_{\rm P}^2} 
  = \frac{\pi \alpha^2}{l_{\rm P}^2},
\label{eq:S_dS}
\end{equation}
gives the logarithm of this number (at the leading order in expansion 
in inverse powers of ${\cal A}/l_{\rm P}^2$), where ${\cal A} = 4\pi 
\alpha^2$ is the area of the de~Sitter horizon~\cite{Nomura:2014yka}. 
In particular, there are exponentially many independent de~Sitter 
{\it vacuum} states---the states that do not have a field or string 
theoretic excitation in the semiclassical background---which all 
represent the same de~Sitter vacuum at the semiclassical level.

The analysis of physics in this de~Sitter vacuum is parallel to that 
on a black hole background in Ref.~\cite{Nomura:2014voa}.  Denoting the 
index representing the exponentially many de~Sitter vacuum states by
\begin{equation}
  k = 1, \cdots, e^{S_0},
\label{eq:k}
\end{equation}
where $|S_0 - S_{\rm GH}| \approx O({\cal A}^q/l_{\rm P}^{2q};\, q<1)$, 
states at late times on this vacuum can be expanded in terms of the 
microstates of the form
\begin{equation}
  \ket{\Psi_{\bar{a} a; k}(\alpha)}.
\label{eq:states}
\end{equation}
Here, $\bar{a}$ and $a$ label excitations of the stretched horizon, 
located at $r = \alpha - O(l_{\rm P}^2/\alpha) \equiv r_{\rm s}$, and 
the interior region, $r < r_{\rm s}$, respectively, where $r$ is the 
static radial coordinate with $r=0$ taken at $p_0$.  Note that excitations 
here are defined as fluctuations with respect to a fixed background, so 
their energies as well as entropies can be either positive or negative, 
although their signs must be the same.  The contribution of the excitations 
to the entropy is subdominant in the $l_{\rm P}^2/{\cal A}$ expansion, 
so that the total entropy of this de~Sitter system (not necessarily 
of the vacuum states) is still given by $S = {\cal A}/4 l_{\rm P}^2$ 
at the leading order.

The indices for the excitations, $\bar{a}$ and $a$, and the vacuum, $k$, 
do not fully ``decouple.''  In particular, operators in the semiclassical 
theory representing modes whose energies defined at $r = 0$ are
\begin{equation}
  \omega \lesssim T_{\rm GH},
\label{eq:IR-modes}
\end{equation}
act nontrivially on both $a$ and $k$ indices, where $T_{\rm GH} = 1/2\pi 
\alpha$ is the Gibbons-Hawking temperature.  This allows us to understand 
the thermal nature of the semiclassical de~Sitter space in the following 
manner.  The fact that all the independent microstates with different 
$k$ lead to the same geometry (within the quantum mechanical uncertainty) 
suggests that the semiclassical picture is obtained after coarse-graining 
the degrees of freedom represented by this index, which we call the 
{\it vacuum degrees of freedom}.  In this picture, the de~Sitter vacuum 
in the semiclassical description is represented by the density matrix
\begin{equation}
  \rho_0(\alpha) = \frac{1}{e^{S_0}} \sum_{k=1}^{e^{S_0}} 
    \ket{\Psi_{\bar{a}=a=0; k}(\alpha)} \bra{\Psi_{\bar{a}=a=0; k}(\alpha)}.
\label{eq:rho_0}
\end{equation}
To obtain the response of this state to the operators in the semiclassical 
theory, we may trace out the subsystem $\bar{C}$ on which they do not act. 
Consistently with our identification of the origin of the Gibbons-Hawking 
entropy, we identify the resulting reduced density matrix as the thermal 
density matrix
\begin{equation}
  \tilde{\rho}_0(\alpha) = {\rm Tr}_{\bar{C}}\, \rho_0(\alpha) 
  \approx \frac{1}{Z} e^{-\frac{H_{\rm sc}(\alpha)}{T_{\rm GH}}},
\label{eq:rho_0-therm}
\end{equation}
where $Z = {\rm Tr}\, e^{-H_{\rm sc}(\alpha)/T_{\rm GH}}$, and 
$H_{\rm sc}(\alpha)$ is the Hamiltonian of the semiclassical theory.

Another manifestation of the non-decoupling nature of the $a$ and $k$ 
indices is that for states having a negative energy excitation, the 
range over which $k$ runs is smaller than that in Eq.~(\ref{eq:k})---this 
is the meaning that a negative energy excitation carries a negative 
entropy.  As discussed in the next section, this fact is important 
in ensuring unitarity in the process in which a physical detector 
held at constant $r$ is excited due to interactions with the de~Sitter 
spacetime.

\section{Vacuum Degrees of Freedom}
\label{sec:vac_dof}

The expression in Eq.~(\ref{eq:rho_0-therm}) implies that the spatial 
distribution of the information in $k$ follows the thermal entropy 
calculated using the local temperature
\begin{equation}
  T(r) = \frac{1}{2\pi\alpha} 
    \frac{1}{\sqrt{1-(\frac{r}{\alpha})^2}}.
\label{eq:T_local}
\end{equation}
Namely, the vacuum degrees of freedom can be viewed (in a given 
quasi-static reference frame) as being distributed according to the 
thermal entropy calculated using $T(r)$ in the semiclassical theory. 
Since the thermal nature is the crucial element in the argument of 
Boltzmann brains, we must ask:\ what is the dynamics of the vacuum 
degrees of freedom?

In Ref.~\cite{Nomura:2014voa}, I have argued, together with Sanches 
and Weinberg, that the thermal nature of semiclassical theory should 
not be taken ``too literally.''  Specifically, it does {\it not} mean 
that the degrees of freedom described within the semiclassical theory 
are actually in thermal equilibrium with local temperature $T(r)$. 
Rather, the thermal nature implies that the vacuum degrees of 
freedom---which are already coarse-grained {\it to obtain} the 
semiclassical theory---interact with the excitations in the semiclassical 
theory---e.g.\ a detector located in de~Sitter space---as if these 
excitations are immersed in the thermal bath of temperature $T(r)$. 
In particular, the dynamics of the vacuum degrees of freedom themselves 
cannot be described by the semiclassical Hamiltonian $H_{\rm sc}(\alpha)$. 
These degrees of freedom are neither matter nor spacetime; they are 
``some stuff'' that reveal either feature of matter or spacetime 
depending on the question one asks---the phenomenon we referred to 
as {\it spacetime-matter duality}.%
\footnote{This situation reminds us of wave-particle duality, which 
 played an important role in early days in the development of quantum 
 mechanics---a quantum object exhibited dual properties of waves 
 and particles, while the ``true'' (quantum) description did not 
 fundamentally rely on either of these classical concepts.}

\begin{figure}[t]
\begin{center}
  \includegraphics[height=10cm]{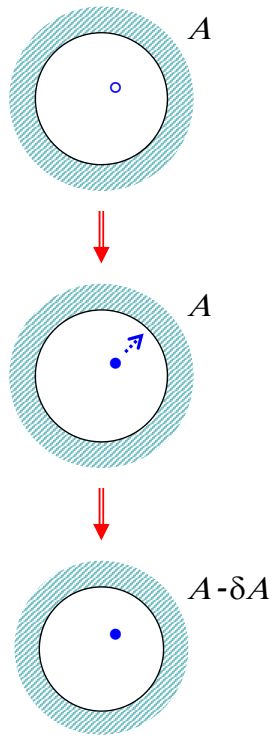}
\end{center}
\caption{A schematic depiction of the process in which a detector 
 located in de~Sitter space interacts with the spacetime; time flows 
 from the top to the bottom.  The detector initially in some state 
 (unfilled dot in the top panel) will react to the local Gibbons-Hawking 
 temperature (filled dot in the middle panel).  This is accompanied 
 by the creation of a negative energy excitation, which propagates 
 to the stretched horizon (dashed arrow in the middle panel).  The 
 background system then eventually relaxes into a new space whose 
 horizon area is smaller than the original one (the bottom panel).}
\label{fig:dS}
\end{figure}

To elucidate this point further, let us consider a physical detector 
held at some constant $r = r_{\rm d}$.  (For $r_{\rm d} \neq 0$, this 
is an accelerating detector.)  The detector then responds {\it as if} 
it is immersed in the thermal bath of temperature $T(r_{\rm d})$, and 
correspondingly extracts information from the vacuum degrees of freedom 
$k$.  It need not, and in fact does not, imply that the semiclassical 
degrees of freedom are actually in thermal equilibrium with the 
temperature in Eq.~(\ref{eq:T_local}).  Due to energy conservation, 
this response is accompanied by a creation of a negative energy 
excitation, which propagates (or free-falls) toward larger $r$ and 
collides with the stretched horizon.  The background spacetime will 
then eventually relax into the one whose horizon area is (slightly) 
smaller, reflecting the existence of the detector with a higher energy. 
This whole process is depicted schematically in Fig.~\ref{fig:dS}. 
Note that because the negative energy excitation has a negative 
entropy, each step in the process can be separately unitary.  For 
more details, see the discussion on the analogous process of black 
hole mining in Ref.~\cite{Nomura:2014voa}.

The second process we consider is the Hawking-Moss 
transition~\cite{Hawking:1981fz} from a de~Sitter vacuum to 
another vacuum.  As discussed, e.g., in Ref.~\cite{Weinberg:2006pc}, 
this transition can be viewed as a thermal process occurring through 
a field climbing up the potential barrier separating the two vacua. 
In our picture, this interpretation becomes valid only in the context 
of a theory built on the daughter vacuum to which the original 
vacuum decays.  In particular, the existence of the transition 
does not imply that the semiclassical degrees of freedom built 
on the original de~Sitter vacuum were actually in thermal equilibrium 
with Eq.~(\ref{eq:T_local}) before the transition.

The two examples above illustrate that the thermal nature of spacetime 
acquires a clear semiclassical interpretation only in the context 
of the vacuum degrees of freedom interacting with other degrees of 
freedom, either semiclassical excitations (e.g.\ when a physical 
detector exists) or another system beyond the one built on the vacuum 
degrees of freedom themselves (e.g.\ when the vacuum decays).  In the 
case of a black hole, the former corresponds to the situation in black 
hole mining, while the latter to spontaneous Hawking emission in which 
the far exterior region (outside the ``zone'') serves as the other 
system~\cite{Nomura:2014voa}.  We note that the picture described here 
does not affect the standard calculation of density fluctuations in an 
inflationary universe~\cite{Hawking:1982cz}, since these fluctuations 
are interpreted in a late-time universe with much smaller vacuum energy.

\section{Boltzmann Brains}
\label{sec:Boltzmann}

Consider a de~Sitter vacuum $J$ with Hubble radius $\alpha_J$.  What 
are the conditions that its decay rate $\Gamma_J$ must satisfy?  If 
the fundamental theory does not have a stable de~Sitter vacuum, as 
suggested by string theory, then the vacuum must decay before the 
Poincar\'{e} recurrence time $t_{{\rm rec},J}$:
\begin{equation}
  \Gamma_J \gtrsim \frac{1}{t_{{\rm rec},J}}.
\label{eq:cond-1}
\end{equation}
Since the Gibbons-Hawking entropy is obtained under the (implicit) 
assumption that the dynamics of the degrees of freedom it represents 
takes a local form, it is appropriate to use for $t_{{\rm rec},J}$ 
the expression for the {\it classical} Poincar\'{e} recurrence time
\begin{equation}
  t_{{\rm rec},J} \sim \alpha_J\, e^{S_{{\rm GH},J}},
\label{eq:t_rec}
\end{equation}
where $S_{{\rm GH},J} = \pi \alpha_J^2/l_{\rm P}^2$.  Note that this 
does not necessarily mean that the dynamics of the vacuum degrees of 
freedom is local in the original de~Sitter space.  It only implies that, 
assuming the vacuum degrees of freedom indeed comprise $O(S_{\rm GH})$ 
quantum degrees of freedom, their dynamics can be organized to take a 
local form in {\it some} (``holographic'') space, i.e.\ the interaction 
Hamiltonian takes a special nearest-neighbor form in this space.%
\footnote{If the dynamics of $O(S_{\rm GH})$ degrees of freedom 
 took a ``fully ergodic'' form in the sense that a generic initial 
 state probes, as time passes, the entire Hilbert space without 
 any ``classicalization,'' then $t_{{\rm rec},J}$ would be given 
 by the {\it quantum} Poincar\'{e} recurrence time, $\alpha_J\, 
 e^{e^{S_{{\rm GH},J}}}$.}

Another requirement for $\Gamma_J$ is that Boltzmann brains do not 
overwhelm ordinary observers in this vacuum, $J$.  This leads to the 
condition~\cite{Nomura:2012zb,Bousso:2006xc}
\begin{equation}
  \Gamma_J \gtrsim \frac{\Gamma_{{\rm BB},J}}{n_J},
\label{eq:cond-2}
\end{equation}
where $n_J$ and $\Gamma_{{\rm BB},J}$ are, respectively, the number 
of ordinary observers and the rate of producing Boltzmann brains, both 
counted with a common rule in the spacetime region causally accessible 
from $p_0$.  Given how the universe enters into $J$, $n_J$ can be computed 
(in principle) using semiclassical theory built on $J$.  The question 
is:\ how to calculate, or estimate, $\Gamma_{{\rm BB},J}$?

Traditionally, $\Gamma_{{\rm BB},J}$ has been estimated using the 
semiclassical theory built on $J$ with the assumption that the degrees 
of freedom {\it in the theory} are in thermal equilibrium with the 
Gibbons-Hawking temperature~\cite{Dyson:2002pf,Bousso:2006xc,%
Freivogel:2008wm,Bousso:2008hz}.  In our picture, however, it is 
the {\it internal dynamics} of the {\it vacuum} degrees of freedom 
that is relevant for the production of Boltzmann brains, which---as 
we have argued in Section~\ref{sec:vac_dof}---cannot be captured by 
the semiclassical Hamiltonian $H_{{\rm sc},J}$.%
\footnote{We assume that the process of ``consciousness'' needed to 
 characterize Boltzmann brains, as well as ordinary observers, is 
 defined by {\it a set of} states spanning the time for the process 
 (not just by an instantaneous state), and thus depends on the 
 Hamiltonian generating the time evolution.}
In fact, we may expect that this dynamics is very different from that 
given by $H_{{\rm sc},J}$.  (Note that if the two were identical, it 
would reintroduce the firewall problem of Ref.~\cite{Almheiri:2012rt}.) 
In particular, we know that $n_J$ is nonvanishing in the vacuum 
representing our universe, but this does not mean that the internal 
dynamics of the corresponding vacuum degrees of freedom (which we do 
not know yet) must produce intellectual observers.  If this dynamics 
does not support any intellectual observer, then the timescale for 
Boltzmann brain production need not be much shorter than the Poincar\'{e} 
recurrence time, i.e.\ the timescale in which the vacuum spontaneously 
creates semiclassical excitations with significant probabilities. 
The decay rate of our universe then need not be much larger 
than $1/t_{{\rm rec},J}$, where $t_{{\rm rec},J}$ is given by 
Eq.~(\ref{eq:t_rec}).

Now, suppose that the fundamental theory has a multitude of vacua, as 
suggested by string theory, and that it leads to the eternally inflating 
multiverse.  Under rather general assumptions about the dynamics of 
the multiverse, the conditions for avoiding Boltzmann brain dominance 
can be written as
\begin{equation}
  \Gamma_J \gtrsim \Gamma_{{\rm BB},J},
\label{eq:cond-3}
\end{equation}
{\it for all de~Sitter vacua} in the theory~\cite{Bousso:2008hz}. 
(The factors $n_J$ do not play a significant role if the probability 
of producing ordinary observers in our universe is not double-exponentially 
suppressed, which seems to be the case.)  In the traditional picture, 
this imposes a strong constraint on the decay rate of {\it any} de~Sitter 
vacuum $J$ supporting intelligent observers.  In particular, it gives 
constraints on the decay rates of all the vacua that are similar to 
our own vacuum, which are much stronger than the ones needed to avoid 
the Poincar\'{e} recurrence.  While there is a suggestion that these 
strong constraints may indeed be satisfied in (at least, a particular 
corner of) the string landscape~\cite{Freivogel:2008wm}, one might 
feel disconcerting that such a fundamental property as the fact that 
we can comprehend the world relies on ``accidental,'' numerical features 
of the theory.  Our picture offers a much simpler possibility:\ the 
dynamics of the vacuum degrees of freedom may simply not support any 
intelligent observers.  If this is the case, then
\begin{equation}
  \Gamma_{{\rm BB},J} \left\{ \begin{array}{ll} 
    \!\!\sim\;\; \frac{1}{\alpha_J} e^{-S_{{\rm GH},J}}
        & \mbox{if the semiclassical theory in $J$ supports observers},\\
    \!\!=\;\; 0 
        & \mbox{otherwise},
    \end{array} \right.
\label{eq:Gamma_BB}
\end{equation}
and the conditions in Eq.~(\ref{eq:cond-3}) can be easily satisfied under 
the assumption in Eqs.~(\ref{eq:cond-1},~\ref{eq:t_rec}).  We may say 
that ``spacetime cannot think.''

We finally note that while some features appearing in the present 
framework look similar to those discussed in Ref.~\cite{Boddy:2014eba}, 
the underlying physical pictures are different, so that the physical 
implications of the two are also different.  In the present picture, 
a semiclassical de~Sitter vacuum is not an exact energy eigenstate and 
is subject to a nontrivial dynamics at the microscopic level (at least 
in cosmological contexts).  The rate of Boltzmann brain production 
in such a vacuum then depends crucially on the (unknown) microscopic 
dynamics of quantum gravity.  This issue was not discussed in 
Ref.~\cite{Boddy:2014eba}.

\section{Summary and Discussion}
\label{sec:discuss}

Assuming that the concept of consciousness is defined physically (and 
that quantum mechanics provides a correct description of nature at the 
fundamental level), the fact that we observe an ordered, comprehensible 
world implies special structures of quantum operators characterizing 
our observations, which act on a Hilbert space in which the state 
representing the universe/multiverse lives.  In particular, in the 
standard time-evolution picture, which arises from focusing on branches 
in the full universe/multiverse state, the decay rate of any de~Sitter 
vacuum that is cosmologically populated must be larger than the 
production rate of Boltzmann brains, Eq.~(\ref{eq:cond-3}).  Under 
certain weak assumptions on the dynamics of the multiverse, satisfying 
this condition is equivalent to explaining the origin of the arrow 
of time we observe in nature.

In this paper we have argued that, in contrast with the traditional view, 
the rate of Boltzmann brain production in a de~Sitter vacuum cannot be 
calculated using the semiclassical theory built on this vacuum.  It is 
determined, instead, by the (unknown) microscopic dynamics {\it of the 
vacuum degrees of freedom}, which is not the same as that of the usual 
semiclassical degrees of freedom despite the fact that they provide the 
origin of the Gibbons-Hawking entropy (the phenomenon referred to as 
spacetime-matter duality in Ref.~\cite{Nomura:2014voa}).  In particular, 
the fact that a physical detector located in the de~Sitter space sees 
a thermal bath of temperature $T(r)$ in Eq.~(\ref{eq:T_local}) does 
{\it not} imply that the semiclassical degrees of freedom, whose dynamics 
is determined by the semiclassical Hamiltonian, are actually in thermal 
equilibrium with temperature $T(r)$.  It only implies the existence of 
{\it some} degrees of freedom---the vacuum degrees of freedom---which 
interact with semiclassical degrees of freedom as if they are a thermal 
bath of temperature $T(r)$.

The picture of the microscopic structure of quantum gravity described 
above offers a new possibility to avoid the dominance of Boltzmann 
brains---the dynamics of the vacuum degrees of freedom may simply not 
support any intelligent observers.  The picture also addresses the 
question~\cite{Page:2005ur} of whether a Minkowski vacuum produces 
Boltzmann brains---it depends on the microscopic dynamics of the vacuum 
degrees of freedom comprising that Minkowski vacuum.  The fact that 
the Hawking temperature is zero in a Minkowski vacuum does not, by 
itself, guarantee the absence of Boltzmann brains; it simply means 
the absence of interactions between the semiclassical and vacuum 
degrees of freedom.  Whether Boltzmann brains are produced in this 
vacuum then depends on the {\it internal} dynamics of the vacuum 
degrees of freedom, which is not yet known.

If the state of the universe/multiverse in fact probes a Minkowski 
vacuum, or stays in a de~Sitter vacuum supporting intelligent observers 
for a very long time, then the consideration in this paper becomes 
relevant.  It is intriguing that such a basic fact that we observe 
an ordered, comprehensible world may deeply rely on the structure 
of quantum gravity at the fundamental level.

\section*{Acknowledgments}

I would like to thank Fabio Sanches and Sean Weinberg for useful 
discussions.  This work was supported in part by the Director, Office 
of Science, Office of High Energy and Nuclear Physics, of the U.S.\ 
Department of Energy under Contract DE-AC02-05CH11231, and in part 
by the National Science Foundation under grant PHY-1214644.

\end{document}